\documentstyle[12pt]{article}

\def\thefootnote{\fnsymbol{footnote}}

\def\bli{\bigg|}
\def\diff{\mbox{d}}
\def\superint {\int\diff^{4}\theta}
\def\bea{\begin{eqnarray}}
\def\eea{\end{eqnarray}}
\def\beq{\begin{equation}}
\def\eeq{\end{equation}}
\def\tF{{\tilde F}}
\def\tZ{{\tilde Z}}
\def\tG{{\tilde G}}

\def\[{\left [}
\def\]{\right ]}
\def\<{\left <}
\def\>{\right >}
\def\({\left (}
\def\){\right )}
\def\lbr{\left\{}
\def\rbr{\right\}}
\def\pp{\partial}

\def\M{\bar{M}}
\def\T{\bar{T}}

\def\R{{\bar{R}}}

\def\Tr{{\rm Tr}}
\def\G{{\cal G}}

\def\L{{\cal L}}
\def\D{{\cal D}}
\def\notD{\not{\hspace{-.05in}\D}}
\def\bl{\bar{\lambda}}

\def\t{\bar{t}}
\def\W{\overline{W}}

\def\hK{\hat{K}}

\def\n{\bar{n}}
\def\m{\bar{m}}

\def\bc{\bar{\chi}}

\def\bc{\bar{\chi}}

\def\bD{\bar{\D}}
\def\da{{\dot\alpha}}

\def\bph{\bar{\phi}}
\def\bF{\bar{F}}
\def\bU{\bar{U}}
\def\bu{\bar{u}}

\def\ww{W^\alpha W_\alpha}
\def\bww{\W_{\da}\W^{\da}}
\def\ll{\lambda\lambda}

\begin{document}

\begin{titlepage} 
\begin{center}
            \hfill    LBNL-43259 \\
            \hfill    UCB-PTH-99/21 \\
   	    \hfill    JHU-TIPAC-99004\\		
            \hfill hep-th/9905122\\
	\hfill May, 1999
\vskip .1in
{\large \bf Gaugino Masses in Modular Invariant
Supergravity}\footnote{This work was supported in part by the
Director, Office of Energy Research, Office of High Energy and Nuclear
Physics, Division of High Energy Physics of the U.S. Department of
Energy under Contract DE-AC03-76SF00098 and in part by the National
Science Foundation under grants PHY-95-14797 and PHY-94-04057.}\\[.1in]

Mary K. Gaillard, Brent Nelson

{\em Department of Physics, University of California, and 

 Theoretical Physics Group, 50A-5101, Lawrence Berkeley National Laboratory, 
      Berkeley, CA 94720, USA}

{\em and}

Yi-Yen Wu 

{\em Department of Physics and Astronomy, Johns Hopkins University,
Baltimore, Maryland 21218, USA }
\end{center}

\begin{abstract}
	We calculate gaugino masses in string-derived models
with hidden-sector gaugino condensation.
The linear multiplet formulation for the dilaton superfield is used to 
implement perturbative modular invariance.
The contribution arising from quantum effects in the
observable sector includes the term recently 
found in generic supergravity models.  A much larger contribution is 
present if matter fields with Standard Model gauge couplings also couple to
the Green-Schwarz counter term.  We comment on the relation of 
our K\"ahler $U(1)$ superspace formalism to other calculations.
\end{abstract}
\end{titlepage}
\newpage

\renewcommand{\thepage}{\roman{page}}
\setcounter{page}{2}
\mbox{ }

\vskip 1in

\begin{center}
{\bf Disclaimer}
\end{center}

\vskip .2in

\begin{scriptsize}
\begin{quotation}
This document was prepared as an account of work sponsored by the United
States Government. While this document is believed to contain correct 
 information, neither the United States Government nor any agency
thereof, nor The Regents of the University of California, nor any of their
employees, makes any warranty, express or implied, or assumes any legal
liability or responsibility for the accuracy, completeness, or usefulness
of any information, apparatus, product, or process disclosed, or represents
that its use would not infringe privately owned rights.  Reference herein
to any specific commercial products process, or service by its trade name,
trademark, manufacturer, or otherwise, does not necessarily constitute or
imply its endorsement, recommendation, or favoring by the United States
Government or any agency thereof, or The Regents of the University of
California.  The views and opinions of authors expressed herein do not
necessarily state or reflect those of the United States Government or any
agency thereof, or The Regents of the University of California.
\end{quotation}
\end{scriptsize}

\vskip 2in

\begin{center}
\begin{small}
{\it Lawrence Berkeley Laboratory is an equal opportunity employer.}
\end{small}
\end{center}

\newpage
\renewcommand{\thepage}{\arabic{page}}
\setcounter{page}{1}
\def\thefootnote{\arabic{footnote}}
\setcounter{footnote}{0}

It was recently pointed out~\cite{rs,hit} that the super-Weyl anomaly
of standard $N=1$
supergravity generates a gaugino mass proportional to the 
beta-function coefficient, which may solve the problem of small
gaugino masses found in certain classes of models.

In this paper we consider a class of string-derived models~\cite{us1}--\cite{us3} 
in which gaugino condensation occurs in a hidden sector with modular invariant 
couplings.  That is, the field theoretic quantum anomaly that breaks 
invariance under the modular transformation (T-duality)
\beq T\to {aT - ib\over icT +d}, \quad ab-cd = 1,\quad a,b,c,d\in Z,\label{tdual}\eeq
is explicitly canceled by a universal Green-Schwarz (GS) counter term together with
model-dependent string threshold corrections.  Gaugino masses in these models were 
found in~\cite{us3} to be suppressed with respect to the gravitino mass--although
not as severely as in some gauge-mediated models~\cite{small}.
However, in~\cite{us3} a contribution was omitted that generates, among others,
the term found in~\cite{rs,hit}.  
In this paper we correct this omission. The additional correction is obtained
by an analysis of the superspace expression for the loop correction, as well as by an
 explicit calculation using component fields and Pauli-Villars regularization.
 In these models the anomaly associated with the K\"ahler 
transformation (\ref{tdual}) is explicitly canceled.  
Because K\"ahler and super-Weyl transformations are intimately 
connected in the K\"ahler
$U(1)$ superspace formalism~\cite{bgg} that we use,
one might expect the mass term found 
in~\cite{rs,hit} to be absent in this class of models. 
However, this term has its origin in the running of the couplings
from the string scale to the condensation scale, and is therefore independent
of the string scale physics.  In addition, we find a contribution
that depends on the unknown couplings of matter fields
in the GS term -- a situation similar to the case for scalar masses
discussed in~\cite{us3}.

In the linear supermultiplet formulation the dilaton $\ell$ is the lowest component of a vector 
superfield $L$ that satisfies the modified linearity condition 
\beq - \(\bD^2 - 8R\)L = \ww, \quad - \(\D^2 - 8\R\)L = \bww, \quad L\bli = \ell,
\label{mod}\eeq
where the superfield $R$ is related to elements of the supervielbein, 
$W^\alpha$ is a Yang-Mills superfield strength, 
and the summation over gauge indices is suppressed.  The Bianchi identity
\beq \(\D^2 - 24\R\)\ww -  \(\bD^2 - 24R\)\bww = {\rm total \; derivative}
\label{bianc}\eeq
follows immediately from (\ref{mod}).  To describe gaugino condensation~\cite{bdqq,bgt},
a vector multiplet $V$ is introduced whose components include those of a linear multiplet
$L$ as well as chiral and anti-chiral superfields $U,\bU$ that are the (anti-)chiral projections
of $V$:
\beq  - \(\bD^2 - 8R\)V = U, \quad - \(\D^2 - 8\R\)V = \bU, \label{cond} \eeq
and are interpreted as condensate superfields for a strongly coupled (confined) 
hidden Yang-Mills sector: $U\simeq (\ww)_h$.  With this construction the superfield
$U$ has the correct K\"ahler $U(1)$ weight as well as the correct constraint, 
that is, the counterpart of the Bianchi identity (\ref{bianc}).
This construction was generalized in~\cite{us2} to the case of several gaugino
condensates, and it was found that the results are dominated by the condensate
of the gauge group $\G_a=\G_+$ with the largest $\beta$-function coefficient $b_a =
b_+$, where 
\bea b_a = {1\over8\pi^2}\(C_a - {1\over3}C_a^M\), \quad C_a^M = \sum_AC^A_a,\eea
with $C_a$ and $C_a^A$ quadratic Casimir operators in 
the adjoint and matter representations, respectively. 
For this reason we include only a single condensate here.  When both the condensate and the 
weakly coupled, unconfined Yang-Mills sectors are included, the linearity condition
takes the form
\bea  - \(\bD^2 - 8R\)V &=& U + \sum_a(\ww)_a, \nonumber \\ - \(\D^2 - 8\R\)V &=&
\bU + \sum_a(\bww)_a, \label{all} \eea

We consider a class of orbifold compactifications with three untwisted moduli chiral superfields
$T^I$ and matter chiral superfields $\Phi^A$. The K\"alher potential is
\beq K = k(V) + \sum_Ig^I + \sum_A e^{q^A_I}|\Phi^A|^2 + O(\Phi^4),\quad g^I =
- \ln(T^I + \T^I), \eeq
where the parameters $q^A_I$ are the modular weights of $\Phi^A$, and
the relevant part of the Lagrangian is 
\bea \L_{eff} = \L_1 + \sum_a\L_a, \quad \L_1 
= \L_{KE} + \L_{GS} + \L_{th} + \L_{VY} + \L_{pot},\eea
where
\bea \L_{KE} = \superint\,E \[-2 + f(V)\], \quad k(V) = \ln\,V + g(V),  \eea
contains the kinetic energy terms for the dilaton, chiral and gravity superfields, as well
as the tree-level Yang-Mills terms. The
functions $f(V),g(V)$ parameterize nonperturbative string effects. They satisfy the conditions
\bea Vg'(V) = f -Vf'(V), \quad g(0) = f(0) = 0,
\eea which ensure that the Einstein term has canonical form~\cite{us1}, and that
they vanish in weak coupling limit: $g^2/2 = \<\ell\> = \<V|\> \to 0$.
The term
\bea \L_{VY} &=& {1\over8}\superint\,{E\over R}U\[b'_+ \ln(e^{-K/2}U)
 + \sum_\alpha b^\alpha \ln\Pi^\alpha\] + {\rm h.c.},\nonumber \\  
b'_a &=& {1\over8\pi^2}\(C_a - C^M_a\),\quad
\sum_\alpha b^\alpha = {1\over12\pi^2}C^M_+,\label{lvy}\eea
is the generalization to supergravity~\cite{tom,bg89} of the 
Veneziano-Yankielowicz superpotential term~\cite{vy}, including~\cite{matter} gauge 
invariant composite matter fields $\Pi^\alpha$, and
\bea \L_{pot} &=& {1\over2}\superint\,{E\over R}e^{K/2}W(\Pi^\alpha,T^I) + {\rm h.c.} \eea
is a superpotential for the matter condensates. In (\ref{lvy}) $C^M_+$ refers to the confined
matter superfields $\Phi^A$ of the strongly coupled sector.  

The operators $\L_a$ are the 
quantum corrections from light field loops to the unconfined Yang-Mills couplings:
\bea \L_a &=& 
{1\over64\pi^2}\int d^4\theta 
{E\over R}\(W_\alpha P_\chi\[f_a(\Box_\chi) - B_a\]W^\alpha\)_a + {\rm h.c.},
\label{la}\eea
where $\Box^{-1}_\chi$ is the chiral superfield propagator~\cite{1001}: 
\beq \<\Box_\chi\> = \<\Box + {1\over2}R\D^2 + O(R\R)\>, \eeq
 in our notation,\footnote{We set the background space-time curvature scalar $r$
to zero throughout this paper.  A term proportional to $r\lambda\lambda$ would result
in a contribution to the gaugino mass through a Weyl rescaling, but we find that such
terms are suppressed by powers of $\mu^{-2}$ or $m^{-2}$ where $m$ is the Pauli-Villars
mass introduced below.}
 and $P_\chi$ is the chiral projection operator:
$P_\chi W^\alpha = W^\alpha$, that reduces
in the flat space limit to $(16\Box)^{-1}\bD^2\D^2$. The function~\cite{gt}
\bea B_a &=& \sum_I\(C - b^I_a\)g^I + \(C^a - C^a_M\)k(V)
+ 2\sum_AC^a_A\ln\(1 + p_AV\),\nonumber \\ 
b^I_a &=& C - C_a + \sum_A\(1 - 2q^A_I\)C^A_a, \quad C = C_{E_8} = 30,\eea
determines the renormalized coupling constant~\cite{us3,tom,kl} $g_a(\mu_s)$ at the 
string scale $\mu_s$:
\bea  g^{-2}_a(\mu_s) &=& \Bigg<{1+f\over2\ell} - b'_ak(\ell) 
+ \sum_A{C_a^A\over8\pi^2}\ln(1+\ell p_A)
\nonumber \\ & & - \sum_I{b^I_a\over16\pi^2}\ln\[|\eta(it^I)|^2(t^I + \t^I)\]\Bigg>, 
\quad \mu_s =\< e^{{1\over2}(k-1)}\>\label{loop}, \eea
and the functions
\beq {1\over 8\pi^2}f_a(\mu^2) = g_a^2(\mu^2) - g_a^2(\mu_s^2)\eeq 
govern the running of the gauge couplings from the string scale to 
the normalization scale $\mu^2 = -<\Box>$.
$\L_a$ and $\L_{VY}$ are anomalous under (\ref{tdual}). 
This anomaly is canceled by two counter terms: the GS term~\cite{gs}
\bea \L_{GS} &=& \superint\,E VV_{GS},\quad b = {C\over8\pi^2} = b_{E_8}, \nonumber \\
V_{GS} &=& b\sum_Ig^I + \sum_A p_Ae^{\sum_Iq^A_Ig^I}|\Phi^A|^2 + O(\Phi^4),\eea
and the term induced by string loop corrections~\cite{dkl}
\beq \L_{th} = - {1\over64\pi^2}\sum_I\superint\,{E\over R}\ln\eta^2(iT^I)\(b_+^IU 
+ \sum_ab_a^I(\ww)_a\) + {\rm h.c.}.\eeq
The parameters $b^I_a$ vanish for orbifold compactifications with no $N=2$ supersymmetry 
sector~\cite{ant}. For $a=+$, the $q^A_I$ are modular weights of the confined matter 
superfields. Note that we have not introduced kinetic terms for the condensate superfield; that
is, we are treating the condensate as static. A dynamical condensate has been studied~\cite{yy} 
in the case of an $E_8$ gauge group, and it was found that the bound state masses
are above the condensation scale; when these states are integrated out the theory reduces
to the static case considered here.

To evaluate the gaugino masses, we set all matter fields to zero in the vacuum:
$<\phi^A> = <\Phi^A|> = 0.$    First recall that ``D-terms'' like $\L_{GS}$ and
$\L_{KE}$ can be cast in the form of ``F-terms'' by integration by parts:
\beq \L = \superint E\phi = - {1\over16}\superint{E\over R}\(\bD^2 - 8R\)\phi + {\rm h.c.}\eeq
These and the remaining ``F-terms'' can be evaluated using the standard construction~\cite{bgg}
\bea \L &=& {1\over2}\superint{E\over R}{\bf r} = \sqrt{g}\(s - r\M\) + {\rm h.c.} + O(\psi_\mu)
, \nonumber \\r &=& {\bf r}\bli, \quad s = - {1\over 4}\D^2{\bf r}\bli,\eea
with
\bea \ww\bli &=& -\lambda\lambda,\quad u = U\bli, \quad M = - 6R\bli = (\M)^*, \quad \ell = V\bli,
\nonumber \\ t^I &=& T^I\bli, \quad F^I = - {1\over 4}\D^2T^I\bli, \quad
F_u =  - {1\over 4}\D^2U\bli,
\nonumber \\ - {1\over 4}\D^2\ww\bli &=& - {1\over2}\(F^2 +iF\cdot\tF\) - 
\M\ll + 2\bl i\notD\lambda + \cdots.\eea
For example,
\bea \L_{GS} &=& -{1\over16}\superint{E\over R}\(\bD^2 - 8R\)VV_{GS} + {\rm h.c.}\nonumber \\ &=&
{1\over16}\superint{E\over R}\(\sum_a(\ww)_aV_{GS} + V\bD^2V_{GS}\) + {\rm h.c.} + 
\cdots \nonumber \\ &=& {b\over8}\sum_I\(\bl i\notD\lambda g^I - \lambda
\lambda F^Ig'^I\) + {\rm h.c.} + \cdots.\eea
The evaluation of the component form of $\L_1$ is rather
involved and has been carried out explicitly in~\cite{us1,us2} neglecting the 
unconfined Yang-Mills fields.  To include the latter terms we need only make the
substitutions 
\bea u &\to& u - \sum_a(\ll)_a,\quad F_u\to F_u - {1\over 4}\sum_a\D^2(\ww)_a\bli
\eea
in those results.
Using the vacuum values found in~\cite{us2}:
\beq \<F^I\> = 0, \quad \<\M\> = {3\over4}\<\(b_+'\bu - 4\W e^{K/2}\)\> = {3\over4}b_+\<\bu\> =
-3m_{\tilde G},\label{vac}\eeq 
the contribution of $\L_1$ to observable sector gaugino masses found in~\cite{us3} is
\beq \L_1 \ni {1\over2g^2_s}\bl i\notD\lambda + {1\over16\ell^2}\(1+\ell g'\)
\(1 + b_+\ell\)\bu\lambda\lambda + {\rm h.c.},\label{lmass1}\eeq
where $m_{\tilde G}$ is the gravitino mass, and
\beq  g_s = \<\sqrt{2\ell\over1 + f}\> \eeq
is the tree-level field theory coupling constant.
The requirement that the vacuum energy vanishes gives the condition~\cite{us2}
\beq \<\(1+\ell g'\)\(1 + b_+\ell\)^2\> = 3\<\ell^2\>b^2_+,\label{vac2}\eeq
so that, taking into account gauge coupling renormalization,
one gets a contribution to the gaugino mass~\cite{us3}
\beq m^{(1)}_a(\mu_c) = - \<{3g_a^2(\mu_c)b_+^2\bu\over8(1 + b_+\ell)}\> = 
{3g_a^2(\mu_c)b_+\over2(1 + b_+\ell)}m_{\tG},\label{mass1} \eeq
where $\mu_c = |u|^{1\over3}$ is the condensation scale in reduced Planck units. A gravitino 
mass in the TeV range requires $b_+\simeq1/30$, so this contribution to the gaugino masses
is quite small, although it is possible that two-loop renormalization effects between
the condensation scale and the weak scale can
bring masses of this order within experimental bounds~\cite{brent}.

To evaluate (\ref{la}) we drop~\cite{hit} terms of order $<R\R/\Box>
= - m_{\tG}^2/4\mu^2$ to obtain
\bea \<[\D^2,f(\Box_\chi)]\>\Phi = \<8\R\[f'(\Box_\chi)\Box + R
\lbr f'(\Box_\chi) + f''(\Box_\chi)\Box\rbr\D^2\]\>\Phi. \eea
where $\Phi$ is chiral; only the first term on the right hand side contributes to 
gaugino masses; this is the contribution found in~\cite{rs,hit}.  
To lowest order in perturbation theory, 
\bea &&f_a(\Box) =  \(3C_a - C_a^M\)\ln\(\Box/\mu_s^2\), \nonumber \\
&&\<[\D^2,f_a(\Box)]\>\lambda_a = 8\<\R\>\(3C_a - C_a^M\)\lambda_a. \eea
In addition we have:
\bea \<\D^2B_a|\> = -\<4\sum_IF^I{\pp\over \pp T^I}B_a\bli
+ \(\bar{U} - 8V\R\){\pp\over \pp V}B_a\bli\> .\eea
Using the vacuum values (\ref{vac}), we obtain
\bea  - {1\over2}m^{(2)}_a(\L_a) &=& {1\over128\pi^2}\<\D^2\[f_a(\Box_\chi) - B_a\]\>
\nonumber \\ &=& -{\bu\over128\pi^2}\Bigg\{(3C_a - C_M)b_+ - (1 + b_+\ell)\times
\nonumber \\ & & 
\[\ell^{-1}(\ell g' + 1)(C_a - C_M) + 2\sum_AC^A_a{p_A\over(1 + \ell p_A)}
\]\Bigg\}, \label{nonloc}\eea
giving for the full contribution at the condensation scale $\mu_c$
\bea &&\L_1 + \sum_a\L_a \ni \sum_a\[{1\over2g^2_a(\mu_c)}\(\bl i\notD\lambda\)_a +
{M_a\over16}\bu(\ll)_a + {\rm h.c.}\], \nonumber \\ M_a &=&   (1 + b_+\ell)
\[\ell^{-2}\(1+\ell g'\)\(1 + b'_a\ell\) + \sum_A{C^A_ap_A\over4\pi^2(1+\ell p_A)}\] -
3b_ab_+.
\label{lmass}\eea
 We now obtain for the gaugino masses
\beq m_a(\mu_c) = {g_a^2(\mu_c)\over2}\[{3b_+(1+b'_a\ell)\over1 + b_+\ell} - 3b_a + 
\sum_A{C^A_ap_A(1+b_+\ell)\over4\pi^2b_+(1+\ell p_A)}\] m_{\tG}.\label{mass} \eeq

Next we explicitly calculate the gaugino masses using
a Pauli-Villars (PV) regularization that has been 
formulated~\cite{mkg} for supergravity Lagrangians with the dilaton in a
chiral multiplet.  For present purposes, we need only consider the
regulation of loops containing gauge-charged fields.
Because the results below depend only on the K\"ahler potential for the PV fields
and their couplings to the GS term, it is straightforward
to transcribe the analysis to the case where the dilaton is described by a
linear supermultiplet.

To regulate loop corrections to the Yang-Mills self-energy, one needs
gauge-charged PV chiral supermultiplets:
$Z^A$ with signature $\eta_A,$ 
that transform under gauge transformations according to representations
$R_A$; $Y_A$ with the same signature that transform according to the conjugate
representation $\R_A$; and $\Phi_\alpha^a$, 
with signature $\eta_\alpha^a,$ that transforms according to the adjoint representation of the gauge group.
In order to cancel the logarithmic divergences in the Yang-Mills self-energy,
the Casimirs $\Tr_{R_A}\(T^aT^b\) = \delta_{ab} C^a_{R_A},$  and the
signatures of the PV fields must satisfy 
\beq \sum_A\eta_A C^a_{R_A} = - C_a^M, \quad 
\sum_\alpha\eta_\alpha^a  =  3.\label{sigs}\eeq
The K\"ahler potential for these fields
takes the form (setting light gauge-charged fields to zero in the background)
\bea K_{PV} &=& \sum_A\[g^Z_A(V)e^{\sum_I\alpha^I_Ag^I}|Z^A|^2 + 
g^Y_A(V)e^{\sum_I\beta^I_Ag^I}|Y_A|^2\] \nonumber \\ & &
 + \sum_{a,\alpha}g_\alpha(V) e^{\alpha_\alpha\sum_Ig^I}|\Phi_\alpha^a|^2
\equiv  \sum_ng_n(V)G_n(T)|\Phi^n|^2,\label{kpv} \eea
The $V$-dependence of $K_{PV}$ requires an additional term $\L_{PV} 
= \int d^4\theta Ef_{PV}(V)$ in the Lagrangian, where $f_{PV}$ is
related to $K_{PV}$ by the differential equation in (10) that
relates $f$ to $g$:
\beq  f_{PV}(V) = \sum_n f_n(V)G_n(T)|\Phi^n|^2, \quad Vg'_n = f_n - Vf'_n,
\quad \Phi^n = Z^A,Y_A,\Phi_\alpha^a.\eeq
  The component Lagrangian can be obtained following the methods
outlined in~\cite{us1}. 
As shown in~\cite{gt}, supersymmetry of the modular anomaly from field theory
quantum corrections imposes constraints on the parameters in (\ref{kpv}):
\bea
1 &=&  \sum_{\alpha}\eta^a_\alpha\alpha_\alpha, \quad \sum_A\eta_A
\(\alpha^I_A + \beta^I_A\)C^a_{R_A} = - \sum_AC^A_aq^A_I,\nonumber \\
0 &=& \sum_A\eta_A\(\ln h^Z_A + \ln h^Y_A\)C^a_{R_A}, \quad
\sum_\alpha \eta^a_\alpha\ln h_\alpha = k, \nonumber \\ 
 h_n(V) &=& f_n(V) + g_n(V).\label{cutoff}\eea
in agreement with the requirements for full one-loop regularization~\cite{mkg}.
Among the fields $Z^A$ there is a subset $\tZ_\alpha^A$ with the same modular weights and
the same gauge couplings as the light fields $\Phi^A$.
If the parameters $p_A$ in the Green-Schwarz term are nonvanishing, we also require
\bea V_{GS}^{PV} &=& \sum_{A,\alpha} p_Ae^{\sum_Iq_I^Ag^I}|\tZ^A_\alpha|^2,\quad  
g^{\tZ}_A = h^{\tZ}_A = 1,\nonumber \\
\alpha^{\tZ^I}_A &=& q_I^A, \quad \sum_\alpha\eta_\alpha^{\tZ^A} 
= -1.\label{tZ}\eea

The potential 
for the PV scalar fields $\phi^n = \phi^a,z^A,y_A$ is
\bea V &=& {1\over16\ell^2}\(\ell g'(\ell) + 1\)\left|u(1+b_+\ell) 
- 4\ell W^{PV}e^{K/2}\right|^2  \nonumber\\
& & - {3\over16}\left|b_+u - 4W^{PV}e^{K/2}\right|^2  
+ \hK_{n\bar m} F^n \bar F^{\bar m} , \label{pvpot}\eea
where
\bea -\bF^{\m} &=& \hK^{n\m}\lbr\frac{u}{4}G_n\bph^n\[(p_n-b_+)h_n + h'_n\(1 + 
\ell b_+\)\]  + e^{K/2}W^{PV}_n\rbr + O(\phi_m^2), 
\nonumber \\ \hK_{n\bar m}  &=& \(\hK^{n\bar m}\)^{-1}
= G_nh_n\left[ 1+ p_n\ell\right] \delta_{n\bar m}, \quad p_{n\ne\tZ} = 0. \eea
The superpotential $W^{PV}$ contains quadratic terms in  the PV fields 
that give their masses. 
These give rise to ``B-terms'' in the potential (\ref{pvpot}) which take the form
\bea V_B &=& {1\over2}\sum_{m,n}\mu_{mn}a_m\phi^m\phi^n + {\rm h.c.}, \quad\mu_{mn} = 
e^{K/2}W^{PV}_{mn}, \nonumber
\\ a_n &=& e^{K/2}{\bu\over4}\lbr b_+\ell + \(1 + b_+\ell\)\[{2p_n\over 1 + p_n\ell} +
2{h'_n\over h_n}  - {1\over\ell}\(\ell g' + 1\)\]\rbr.\eea
The first term in this expression is independent of the PV 
K\"ahler potential -- {\it i.e.}, of the effective cut-offs -- 
and of the details of the supersymmetry breaking mechanism.
It is precisely the contribution found in \cite{rs,hit}. Here it arises from the presence of 
the K\"ahler potential in the condensate Lagrangian (\ref{lvy}) as dictated
by local supersymmetry. As noted previously~\cite{bgi}, the structure of this term embeds the
evolution of the gauge coupling constant from the string scale to the condensation scale.
The PV Lagrangian also contains the terms (in four-component spinor notation)
\bea\L_{PV}&\ni& \sum_n\hK_{n\n}\[- \pp_a\bph^{\n}\pp^a\phi^n + i\bc_L^{\n}\notD\chi^n_L
+ \sqrt{2}i\(\bl^a_R(\bph^{\n}T_a\chi_L^n) + {\rm h.c.}\)\] \nonumber \\  & &
- \sum_{n,m}
\[\mu^2_{mn}|\phi^n|^2 +  {1\over2}\mu_{mn}\(\bc^n_R\chi^m_L + {\rm h.c.}\)\],\eea
In terms of the normalized fields $\Phi_r^n = (\hK^{1\over2})_{n\n}\Phi^n$,
the interaction terms are
\bea \L_{Yuk} &=& \sqrt{2}i\sum_n\bl^a_R(\bph_r^{\n}T_a\chi_{rL}^n) + {\rm h.c.}, \quad
V =  {1\over2}\sum_{m,n}m_ma_m\phi_r^m\phi_r^n + {\rm h.c.},  \nonumber \\ 
m^2_n &=& \hK^{n\n}\hK^{m\m}\mu^2_{mn} ,\eea
where $m_n$ is the mass of $\Phi_r^n$.
The Feynman amplitude  $F_a = -i<\lambda^a_R|\L_{eff}|\lambda^a_L>$ for 
$\lambda_L \to \chi^n_L + \bph^{\n} \to\phi^m + \bc_R^{\m} \to \lambda_R$
gives a contribution
\bea \L_{eff} &\ni& - {1\over2}m_a\bl^a_R\lambda^a_L + {\rm h.c.} = 
{i\over2}F_a\bl^a_R\lambda^a_L + {\rm h.c.}, \nonumber \\
- {1\over2}m_a^{(2)} &=& -i\sum_n\eta_n C^a_n\int{d^4p\over(2\pi)^4}
{a_nm^2_n\over(p^2 - m_n)^3} =- \sum_n\eta_n {C^a_na_n\over32\pi^2},\eea
 Using the constraints (\ref{sigs}), (\ref{cutoff}) 
and (\ref{tZ}), this reduces to the result found in (\ref{nonloc}).

Aside from the renormalization of the coupling constants, 
there are three contributions from $\L_a$ in (\ref{mass}). 
The first, proportional to
$b'_a,$ gives a negligible correction to (\ref{mass1}).  
The second term modifies the result (\ref{mass1}) of~\cite{us3} by a
factor (neglecting $b_+\ell\sim .03$)
\beq \eta_a \simeq (1 - b_a/b_+) \simeq [0.6,1.1,1.8] \quad {\rm for}
\quad \G_a = \[SU(3),SU(2),U(1)\],\label{pa0}\eeq
if we assume just the MSSM contribution to the $\beta$-functions.
The third term depends on the unknown parameters $p_A$.  It was found in~\cite{us3}
that the squark, slepton and Higgs masses $m_s$ also depend on these parameters.  
If the matter fields decouple from the GS term, one has
\beq p_A = 0,\quad m_s = m_{\tG}, \eeq
and the full correction to the gaugino masses is given by (\ref{pa0}).
If the GS term is proportional to K\"ahler potential we get
\beq p_A = b,\quad m_s \approx 10m_{\tG}. \eeq
Analyses of dynamical symmetry breaking in the MSSM favor smaller masses
for at least the stop and Higgs particles.
Another possibility is that the GS term depends only on the metric of
the compact 6-manifold, in which case it couples only to untwisted fields:
\beq p_A^{untw} = b, \quad m_s^{untw} \approx 10m_{\tG}, \quad p_A^{tw} = 0,
\quad m_s^{tw} = 0,\eeq
resulting in a mass hierarchy among generations as has been proposed by some 
authors~\cite{cohen}.  For a single Standard Model generation, $\sum_AC^A_a = 2$, so
in this scenario, with $n$ untwisted generations, 
the last term in (\ref{mass}) dominates, and one gets the following
gaugino masses at the condensation scale:
\beq m_a(\mu_c) \approx -{n\alpha_a(\mu_c)b(1 + \ell b_+)\over 
\pi b_+(1+\ell b)}m_{\tG}
\approx -{10n\alpha_a(\mu_c)\over\pi} m_{\tG} .\label{mass0} \eeq

Finally, we address the generality of the result (\ref{lmass}) in the broader
context of the class of string-derived models that we are considering.  It would 
be modified if modular invariance is broken by string nonperturbative effects, such as
a moduli-dependence of the functions $g(V),f(V)$ as was found for a particular
orbifold~\cite{eva}.  Modular invariance of the effective Lagrangian for the condensate
ensures that the moduli are stabilized at one of the two self-dual points
in the fundamental domain: $t^I = 1,\;e^{i\pi/6}$.  Together with the condition that the vacuum
energy vanishes, this assures that their F-components vanish in the vacuum: $<F^I>=0$.
The potential for the moduli and the dilaton~\cite{us2} is
\bea 16\ell^2 V &=& \( 1+\ell {dg \over d\ell} \)\left|\(1+b_+\ell\)u_+\right|^2
\nonumber \\ & & - 3\ell^2\left|b_+u_+\right|^2
+ 4\ell^2\(1+b\ell\)\sum_I\left|{F^I\over{\rm Re}t^I}\right|^2, \label{pot}\eea
so the condition (\ref{vac2}), responsible for the suppression of gaugino masses if
$p_A= 0$, holds only to corrections of order $(<F^I>)^2$.
If $<F^I>\ne0$, there are additional corrections to gaugino masses from:
\beq \L_{th} + \L_{GS} + \sum_a\L_a \ni - \sum_{a,I}{b^I_a\over32\pi^2}F^I
\[{\eta'(t^I)\over\eta(t^I)} - {1\over4{\rm Re}t^I}\].\eeq
Since the term in brackets vanishes at the self-dual points, both corrections
are of order $(<F^I>)^2$, and will be small if the moduli are stabilized near
the self-dual points.  Moreover, the second contribution is absent in the $Z_3,Z_7$
orbifolds that appear promising for model building~\cite{iban}.
The condition (\ref{vac2}) provides
several phenomenologically desirable features~\cite{us3} of our model, namely
moduli masses much larger than the gravitino mass:
\beq m_{\ell} \approx 50m_{t^I }\approx 10^3 m_{\tG}, \eeq
and a suppression of the axion decay constant by a factor of about 50 with
respect to earlier estimates~\cite{bd1}.  Moreover, the result that $<F^I>=0$ 
avoids a potential source of unwanted flavor-changing neutral currents.
A detailed analysis of the phenomenology of this 
class of models will be given elsewhere~\cite{brent}.

In concluding, we wish to emphasize that the gaugino mass contribution
equal to $m_0 =\beta(g^2)m_{\tG}/2g^2$ is a model independent result in our 
K\"ahler $U(1)$ superspace formalism, in agreement with the assertion made
in~\cite{hit}.  In the formalism of~\cite{rs}, the auxiliary field of the supergravity 
multiplet differs from the field $M$ used here by a Weyl rotation that depends on
the K\"ahler potential. As a result, the analogous term that they find is not
model independent; for example they get no contribution in no-scale models~\cite{lisa}.
In our formalism, $m_0$ in this case is exactly canceled by the contribution
from $B_a$ in (\ref{la}). Consider for example the simplest no-scale model 
with K\"ahler potential 
  \beq K = - 3\ln(T + \T - \sum_A|\Phi^A|^2), \quad <\phi^A> = 0, \quad
<W>\ne 0 ,\eeq
{\it e.g.} $W = \Phi^3 +$ constant. In this model (with no cancellation of the
modular anomaly) 
\beq B_a = \(C_a - {1\over3}C_a^M\)K, \quad 
\<\D^2B_a|\> = -\<4F^T{\pp\over \pp T}B_a\> \bli, \eeq
the vacuum values satisfy 
\beq F^TK_T = -3e^{K/2}W, \quad R| = {1\over2}e^{K/2}W, \eeq
and we obtain 
\bea  \<\D^2\[f_a(\Box_\chi) - B_a\]\> = 0, \eea
in agreement with~\cite{lisa}.  We find the same cancellation in a PV calculation
 for this model.

\vskip .5cm
\noindent {\bf Acknowledgements}
\vskip .5cm
We thank Lisa Randall and Tom Taylor
for discussions.
  This work was supported in part by the Director, Office 
of Energy Research, Office of High Energy and Nuclear Physics, Division of 
High Energy Physics of the U.S. Department of Energy under Contract 
DE-AC03-76SF00098 and in part by the National Science Foundation under 
grants PHY-95-14797 and PHY-94-04057.

\end{document}